\documentclass[aps,plapplied,twocolumn,psfig,showpacs]{revtex4-1}
\usepackage{amssymb}
\usepackage{amsmath}
\usepackage{graphicx}
\usepackage{color}
\usepackage{ulem}
\makeatletter
\RequirePackage[hyperindex,colorlinks,bookmarksnumbered,plainpages=true]{hyperref}
\hypersetup{colorlinks,linkcolor=blue,urlcolor=blue,citecolor=blue}

\begin{document}

\title{A spincaloritronic battery}

\author{Xiao-Qin Yu$^{1}$}

\author{Zhen-Gang Zhu$^{1,2}$}
\email{zgzhu@ucas.ac.cn}

\author{Gang Su$^{2,3}$}
\email{gsu@ucas.ac.cn}

\author{A. -P. Jauho$^{4}$}
\email{Antti-Pekka.Jauho@nanotech.dtu.dk}

\affiliation{$^{1}$School of Electronic, Electrical and Communication Engineering, University of Chinese Academy of
Sciences, Beijing 100049, China. \\
$^{2}$ Theoretical Condensed Matter Physics and Computational Materials Physics Laboratory, College of Physical Sciences, University of Chinese Academy of Sciences, Beijing 100049, China.\\
$^{3}$ Kavli Institute of Theoretical Sciences, University of Chinese Academy of Sciences, Beijing 100049, China.\\
$^{4}$Center for Nanostructured Graphene (CNG), DTU Nanotech, Department of Micro- and Nanotechnology, Technical University of Denmark, DK-2800 Kgs. Lyngby, Denmark.}

\begin{abstract}
The thermoelectric performance of a topological energy converter is analyzed. The H-shaped device is based on a combination of transverse topological effects involving the spin: the inverse spin Hall effect and the spin Nernst effect. The device can convert a temperature drop in one arm into an electric power output in the other arm. Analytical expressions for the output voltage, the figure-of-merit (ZT) and energy converting efficiency are reported.
We show that the output voltage and the ZT can be tuned by the geometry of the device and the physical properties of the material. Importantly, contrary to a conventional thermoelectric device,
 here a low electric conductivity may in fact enhance the ZT value, thereby opening a path to new strategies in optimizing the figure-of-merit.

\end{abstract}

\maketitle
\section{Introduction}
\label{Introduction}
Conventional thermoelectric (TE) energy converters can be used for recycling waste heat through the Seebeck effect converting the heat current into electric power, or, reversely, be used for TE cooling through the Peltier effect \cite{A.Majumdar,J.C.Peltier}.
The efficiency of TE can be characterized by the dimensionless figure of merit \cite{D.Rowe} $ZT=\frac{S^2\sigma T}{\kappa}$, where $S$ is the Seebeck coefficient, $T$ indicates absolute temperature and $\sigma (\kappa)$ is the electrical (the thermal) conductivity. $\kappa$ has contributions from both electrons and phonons. To optimize the efficiency, $S$ and $\sigma$ should be maximized and $\kappa$ has to be minimized.
However, $\sigma$ usually has a similar dependence on external parameters as $\kappa$. For example, decreasing disorder leads to a larger electrical conductivity, but also $\kappa$ tends to increase at the same time. 
%
Increasing $\sigma$ by a higher charge carrier concentration is usually counteracted by a decreasing Seebeck coefficient $S$.
The conventional strategies to optimize the ZT are based on an attempt to control the electrical conductivity and thermal conductivity separately: one tries to find a material in which electrical conductivity is high but the thermal conductivity (mostly due to phonons) is low.
Owing to the mutual interdependence of the three coefficients ($S, \sigma, \kappa$) it is a daunting challenge to achieve simultaneous optimization in a single material \cite{S.R.Boona}.
In the last twenty years, strategies have been focused on breaking this entanglement \cite{J.P.Heremans}, giving a doubling of the efficiency of the laboratory materials. By careful nanoengineering it is possible to design devices which have a high electrical conductance and a low thermal conductance (see, e.g., Ref. [\onlinecite{MarkussenPRL}]), but the scalability of these devices is challenging.  In spite of the progress, the efficiency of TE devices still remains too low for wide-spread applications.

Spin caloritronics \cite{S.R.Boona,G.E,A.D.Avery,S.Y.Huang,K.Uchida,J.Xiao,H.Adachi}, which is an extension and combination of spintronics and the conventional thermoelectrics, has recently emerged as a new research area. 
Here, a particular aim is paid to the interplay between a temperature gradient and spins, 
and new effects have been discovered which provide a promising platform for improving the thermoelectric performance.
%
Energy converters based on  spin caloritronics have been devised and have, conceptually, advantages over the conventional TE devices. The spins, which behave essentially as an angular momentum, can be manipulated or affected by external magnetic field, ferromagnetic materials, and spin-orbit coupling (SOC). The heat, on the other hand, is mainly carried by phonons which do not carry angular momentum. Therefore, the main two components of spin caloritronics can in principle be controlled independently. This is a great advantage and may lead to high efficiencies for an appropriately designed energy converter.

\begin{figure}[tb]
\centering
\includegraphics[width=0.9\linewidth]{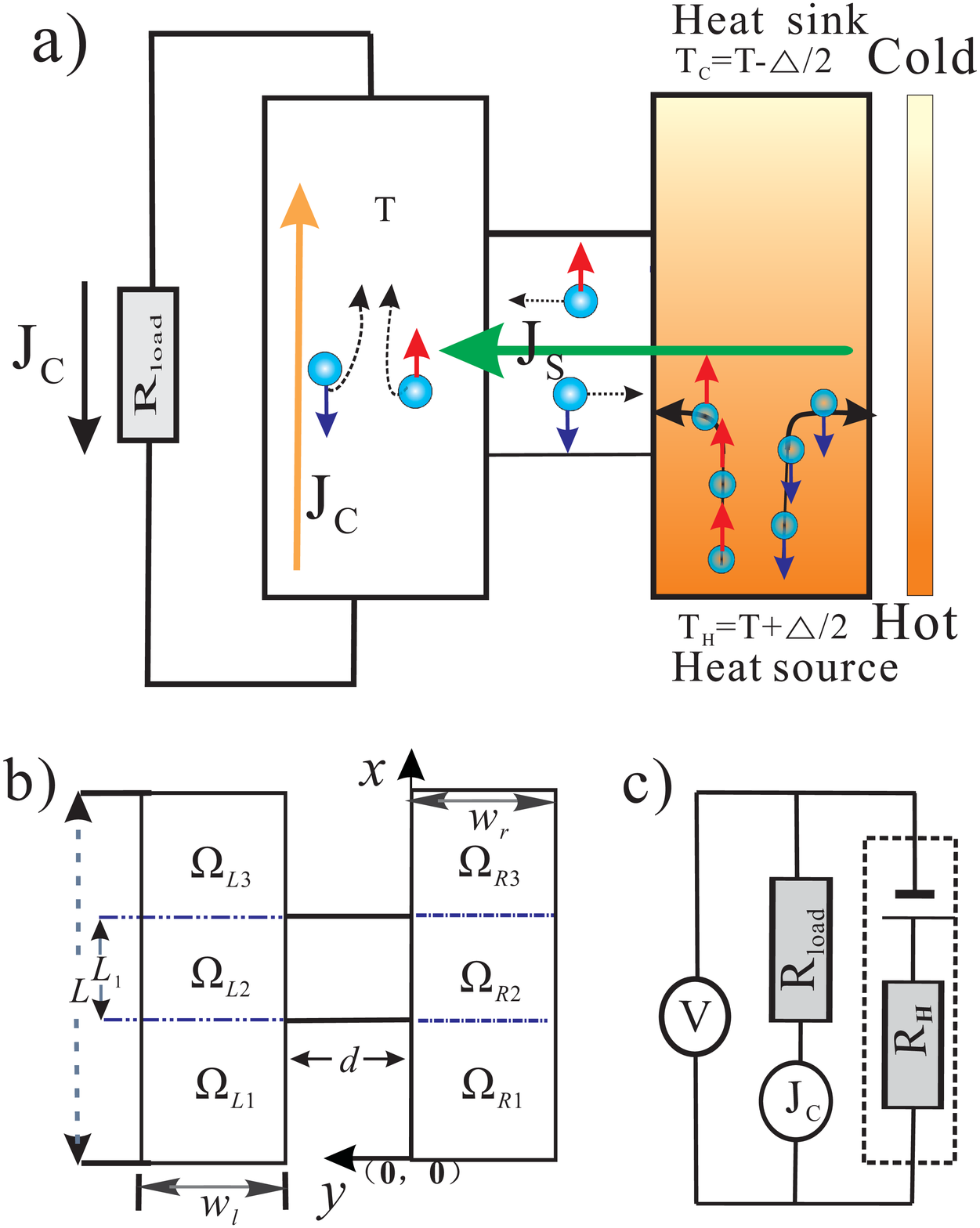}
\caption{(a) A schematic of the spin Nernst effect generator (H-shaped device) based on the ISHE. The spin current $J_{s}$, generated in the right arm by a temperature gradient, is injected into the left arm through a horizontal bridge and then converted into a charge current $J_{c}$ in the $x$-direction by ISHE. (b) The effective computational model. (c) The equivalent circuit for the SNE-based generator: {{the emf generated by the combination of ISHE and SNE in the device is connected to a load resistance $R_{\text{load}}$.  The emf  is equivalent to a battery  with output voltage $\frac{A_{\text{H}}}{G_{\text{H}}}\Delta T$ and internal resistance $R_{\text{H}}=1/G_{\text{H}}$. Here $G_{\text{H}}$ and $A_{\text{H}}$ are the charge conductance and the Nernst conductance of the system given in Eq. (\ref{effective_para}).}}}
\label{1}
\end{figure}

Spin Seebeck effect (SSE)  has earlier been investigated as the driving mechanism in an energy converter \cite{Tianjun Liao,A.B.Cahaya}. 
In 2015, we studied a new effect, 
spin Nernst effect (SNE), and proposed an H-shape device [Fig. \ref{1}(a)] based on monolayer Group-VI dichalcogenides (TMDCs) \cite{Xiao-Q Yu}, to generate pure spin currents.
Due to the SOC in the material and the SNE, a pure transverse spin current can be produced when applying a temperature gradient in the right arm of the device. The spin current can be injected to the left arm through the horizontal bridge. The injected spin current can be converted into a voltage drop along the left arm due to the inverse spin Hall effect (ISHE).
We showed that the voltage drop can be expressed as $\Delta V_{\text{ISHE}}=-\frac{\sigma_{\text{SH}}}{\sigma^{2}}(\frac{2e}{\hbar})\alpha^{\text{spin}}_{n}\Delta T$, where $\sigma_{\text{SH}}$ is spin Hall conductivity.
In this letter, we show that this device can also function as a two-dimensional (2D) thermal battery, where the temperature difference is converted into an electrical power output. In contrast to the conventional TE devices, the mechanisms involved here are two spin-dependent effects, i.e. SNE and ISHE, rather than the conventional Seebeck effect,
 We evaluate the expected device performance, the energy converting efficiency, and the figure-of-merit ZT.  We show that the output voltage and the ZT can be tuned by the geometrical shape and material parameters. We believe that this flexibility in controlling the ZT can be utilized in realistic applications.

\section{system and computational details } \label{scd}
For a temperature gradient along the right arm (x-direction in Fig. \ref{1} (a)), the spin current density $j^{s}_{y}$ along the y-direction and the charge (heat) current density $j^{c}_{x}$ ($j^{Q}_{x}$) along the x direction in the right arm are given in linear response as \cite{G.E,A. B.,M. Hatami, M. Johnson1987,Y.-T.Chen}
\begin{equation}
\begin{aligned}
\begin{bmatrix}
\begin{smallmatrix}
j_{x}^{c} \\
\frac{2e}{\hbar }j_{y}^{s} \\
j_{x}^{Q}%
\end{smallmatrix}
\end{bmatrix}
=
\begin{bmatrix}
\begin{smallmatrix}
\sigma _{r} & \theta _{\text{sHr}}\sigma _{r} & S_{r}\sigma _{r} \\
-\theta _{\text{sHr}}\sigma _{r} & \sigma _{r} & \frac{2e}{\hbar}\alpha _{xy}^{s} \\
S_{r}\sigma _{r}T & -\frac{2e}{\hbar}\alpha _{xy}^{s}T & \kappa_{r}+S_{r}^{2}\sigma _{r}T%
\end{smallmatrix}
\end{bmatrix}
\begin{bmatrix}
\begin{smallmatrix}
-\partial _{x}\mu^{c}_{r}/e \\
-\partial _{y}\mu^{s}_{r}/2e \\
-\partial _{x}T%
\end{smallmatrix}
\end{bmatrix},
\end{aligned}
\label{mt-R-Transport}
\end{equation}
where the subscript ``\textit{r}" refers to the right arm, $\theta_{\text{sHr}}=\sigma_\text{sHr}/\sigma_{r}$ is spin Hall angle, $\sigma_{\text{sHr}}$ is the spin Hall conductivity. $\kappa_{r}$, $S_{r}$ and $\alpha _{xy}^{s}$ are the thermal conductivity, Seebeck coefficient and spin Nernst coefficient, respectively.
In an open circuit, there is no charge current density in the $x$-direction, i.e. $j^{c}_{x}=0$. Therefore, the  electrochemical potential difference $\partial _{x}\mu^{c}_{r}$ is determined by the spin electrochemical potential difference $\partial _{y}\mu^{s}_{r}$ and the temperature gradient $\partial _{x}T$, leading to
\begin{equation}
\begin{aligned}
\begin{bmatrix}
\begin{smallmatrix}
\frac{2e}{\hbar }j_{y}^{s} \\
j_{x}^{Q}%
\end{smallmatrix}
\end{bmatrix}
=
\begin{bmatrix}
\begin{smallmatrix}
\theta _{\text{sHr}}^{2}\sigma _{r}+\sigma _{r} & \theta _{\text{sHr}}\sigma_{r}S_{r}+\frac{2e}{\hbar}\alpha _{xy}^{s} \\
-(\theta _{\text{sHr}}\sigma _{r}S_{r}+\frac{2e}{\hbar}\alpha _{xy}^{s})T & \kappa_{r}%
\end{smallmatrix}
\end{bmatrix}
\begin{bmatrix}
\begin{smallmatrix}
-\frac{\partial _{y}\mu^{s}_{r}}{2e} \\
-\partial _{x}T%
\end{smallmatrix}
\end{bmatrix},
\end{aligned}
\label{msc-current}
\end{equation}
The spin electrochemical potential $\mu_{r}^s$ is determined by the spin-diffusion equation \cite{P.C.Van,M. Johnson1988}
$\nabla^{2}{\mu}_{r}^{s}=\frac{{\mu}^{s}_{r}}{\lambda^{2}_{r}}$,
where $\lambda_{r}=\sqrt{D_{r}\tau_{r,sf}}$ is spin-diffusion length, $\tau_{r,sf}$ is spin-flip relaxation time \cite{P.C.Van}, and $D_{r}=\mu m^{*}v_{F}^{2}/{2}$ is charge diffusion constant determined by mobility $\mu$, the effective mass $m^{*}$, and the Fermi velocity $v_\text{F}\simeq 5.336\times10^{5} m/s$. The spin-flip relaxation time  in MoS$_{2}$ is found to be larger than nanoseconds ($10$ ns $\sim 100$ ns) from both theory \cite{H.Ochoa2014} and experiments \cite{T.Cao,K.F.Mak,H.Zeng}. We use $\mu=400 cm^{2}V^{-1}s^{-1}$ \cite{K.Kaasbjerg} and $m^{*}=0.54 m$ \cite{H.Ochoa} for the hole. Thus, the spin-diffusion length of monolayer MoS$_{2}$ is found to be in the range of $6 \mu m\sim 60 \mu m$. Since $s_{z}$ is a good quantum number \cite{Xiao}, a relative longer spin relaxation length can be expected coinciding with the experimental observations.

As shown in Fig. \ref{1}(b), we divide the right (left) arm into three regions.
Owing to different boundary conditions along the y-direction for  regions $\Omega_{R2}$ and $\Omega_{R1}(\Omega_{R3})$, the temperature gradient in each region instead of the entire right arm is assumed uniform in linear response regime.
The total temperature difference between the ends of the right arm is
$\Delta T=\frac{L-L_{1}}{2}\left(\partial _{x1}T+\partial _{x3}T\right)+L_{1}\partial_{x2}T$, where $\partial _{x1}T$ is derived to be the same to $\partial _{x3}T$ (see  Appendix \ref{app-ltp} for a detailed discussion).
For fixed boundaries in the open circuit case, the spin current flowing in one direction will be balanced by a backflow of spin current in the opposite direction, which leads to zero spin current and spin accumulation at these boundaries. The heat current $J^{\text{Q}}_{x}=\int^{0}_{-w_{r}}j^{\text{Q}}_{x}dy$ is uniform in the entire right arm. Thus, the boundary conditions are
\begin{equation}
\left\{
\begin{array}{lll}
j^{s}_{yi}(y=-w_{r})&=0,  & i=1,2,3, \\
j^{s}_{y2}(y=0)&=j^{s}_{yb}, & \\
j^{s}_{yj}(y=0)&=0,   & i=1,3, \\
J^{Q}_{x1}&=J^{Q}_{x2} &=J^{Q}_{x3},
\end{array}
\right.
\end{equation}
where $j^{s}_{yb}$ is the spin current density in the bridge region, and will be determined below.  The bridge is assumed to be shorter than the spin flip length so that the spin current density can be viewed as spatially independent. With these conditions
the spin accumulation $\mu^{s}_{ri}$ and the temperature gradients $\partial_{xi}T$ in each region are linear functions of the temperature difference $\Delta T$  and the spin current $j^{s}_{y\text{b}}$ in the bridge (see Appendix \ref{app-ltp} for a detailed discussion).
The heat current becomes
\begin{equation}
\begin{aligned}
J_{x}^{Q}&=\left(-\kappa_{r}w_{r}+2\xi_{r}\zeta _{r} \tanh\frac{w_{r}}{2\lambda _{r}}\right)\frac{\Delta T}{L} \\
&+\frac{4e^{2}}{\hbar }\frac{L_{1}\zeta _{r}\lambda _{r}\tanh\frac{w_{r}}{2\lambda _{r}}}{L\Theta\sigma _{r}}j_{y\text{b}}^{s},
\end{aligned}
\label{mheat-current1}
\end{equation}
where $\Theta=\theta _{\text{sHr}}^{2}+1$,  $\zeta_{r}=-\frac{\left(\theta _{\text{sHr}}\sigma _{r}S_{r}+\frac{2e}{\hbar}\alpha _{xy}^{s}\right)T}{2e}$, and $\xi_{r}=\frac{\left(\theta _{\text{sHr}}\sigma _{r}S_{r}+\frac{2e}{\hbar}\alpha_{xy}^{s}\right)2\lambda _{r}e}{\Theta\sigma _{r}}$. 

When a spin current is injected into the left arm through the bridge, a charge current $j^{c}_{x}$ is induced along the $x$-direction owing to the ISHE, which in turn reduces the spin current $j^{s}_{y}$ due to the spin Hall effect (SHE). In linear response
\begin{equation}
\left(
\begin{array}{c}
j_{x}^{c} \\
\frac{2e}{\hbar }j_{y}^{s}%
\end{array}%
\right) =\sigma _{l}\left(
\begin{array}{cc}
1 & \theta _{\text{sHl}} \\
-\theta _{\text{sHl}} & 1%
\end{array}%
\right) \left(
\begin{array}{c}
-\partial _{x}\mu^{c}_{l}/e \\
-\partial _{y}\mu^{s}_{l}/2e%
\end{array}%
\right),
\label{mmatrix_left}
\end{equation}
where  $\sigma_{l}$ is the electrical conductivity of the left arm, $\mu^{c}_{l}=\left(\mu_{\uparrow l}+\mu_{\downarrow l}\right)/2$ is the electrochemical potential, and $\mu_{l}^{s}$ means the spin electrochemical potential of the left arm.
In linear response the induced voltage drop in each region can be assumed to be uniform, which yields
$\Delta V=\frac{L_{1}-L}{2e}\left(\partial _{x}\mu _{l1}^{c}+\partial _{x}\mu_{_{l3}}^{c}\right)-\frac{L_{1}}{e}\partial _{x}\mu_{l2}^{c}$,
where $\bigtriangleup V=V|_{x=0}-V|_{x=L}$.
Analogously, the spin accumulation $\mu^{s}_{li}$ also obeys the spin diffusion equation, i.e. 
$\nabla^{2}\mu^{s}_{l}=\mu^{s}_{l}/\lambda^{2}_{l}$ where $\lambda^{2}_{l}$ is the spin diffusion length of the left arm. %
By using the boundary condition $j_{y}^{s}(y=d+w_{l})=0$ (all regions $\Omega _{l1},\Omega _{l2},\Omega_{l3}$), $j_{y}^{s}(y=d)=0$ (regions $\Omega _{l1}$ and $\Omega_{l3}$), $j_{y}^{s}(y=d)=j_{y\text{b}}^{s}$ (region $\Omega _{l2}$) and the uniform charge current $J^{c}_{x}=\int^{w_{l}+d}_{d}j^{c}_{x}dy$ in the entire left arm, $\mu_{l2}^{s}$/ $\partial _{x}\mu _{li}^{c}$ can be expressed as linear functions of $\Delta V$ and $j_{y\text{b}}^{s}$. The relation between the charge current $J^{c}_{x}$ and the voltage drop along the left arm becomes (details can be found in Appendix \ref{app-lrr})

%
%
\begin{eqnarray}
J_{x}^{c} &=& \int^{w_{l}+d}_{d}j^{c}_{x}dy
= \frac{\Delta V\sigma _{l}}{L}\left(w_{l}+2\theta _{\text{sHl}}^{2}\lambda _{l} \tanh{\frac{w_{l}}{2\lambda _{l}}}\right) \nonumber\\
&\quad&+ \frac{L_{1}}{L}\theta _{\text{sHl}}\lambda _{l}\tanh\left(\frac{w_{l}}{2\lambda _{l}}\right)\frac{2e}{\hbar }j_{y\text{b}}^{s}.
\label{mL_charge_current}
\end{eqnarray}
To obtain an optimal output, spin coherence should be preserved in the bridge. The SOC is the main source of spin relaxation in a material. Nevertheless the $s_{z}$ is a good quantum number in the TMDCs. In addition, owing to the strong spin and valley coupling at the valence-band edges, only atomic scale magnetic scatters would lead to spin flip \cite{Xiao}. In the case of a short bridge operating in the ballistic regime the spins are expected to be conserved.
We also assume that the spin diffusion length is larger than the length of the bridge such that there is no spin accumulation in the bridge,
$\mu _{s}|_{y=0}=\mu _{s}|_{y=d}$.

With known $\mu _{s}|_{y=0}$ ($\mu _{s}|_{y=d}$), the spin current $j_{y\text{b}}^{s}$ can be determined as a function of temperature gradient $\triangle T$ of the right arm and the voltage drop $\triangle V$ generated in the left arm (see Eq. (\ref{spin-birdge})). Then, the relation between various currents and effective forces can be summarized as
\begin{figure}[tb]
\centering
\includegraphics[width=1.0\linewidth]{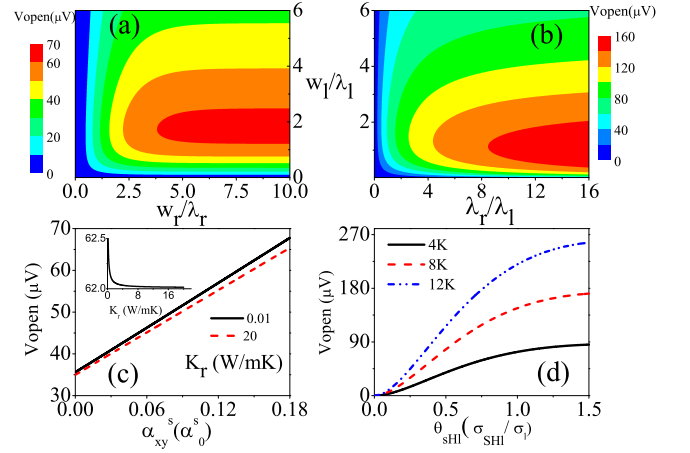}
\caption{ (a) The voltage drop $\text{V}_{\text{open}}$ as a function of $w_{l}/\lambda_{l}$ and $w_{r}/\lambda_{r}$. (b) The voltage drop $\text{V}_{\text{open}}$ as a function of $w_{l}\lambda_{l}$ and $\lambda_{r}/\lambda_{l}$. (c) $\text{V}_{\text{open}}$ versus spin Nernst coefficient $\alpha_{xy}^{s}$ at two different thermal conductivities. (Insert: $\text{V}_{\text{open}}$ versus thermal conductivity). (d) $\text{V}_{\text{open}}$ versus the spin Hall angle $\theta_{\text{sHl}}$ of the left arm at different temperature differences at the two ends of the right arm.  Here $\theta_{\text{sHr}}=0.83, S_{r}=250 \mu\text{VK}$ \cite{Krishnendu}, $\frac{\text{L}_{1}}{\text{L}}=0.5$, T=300\text{K} and $\Delta\text{T}=4K$. Parameters $w_{r}/\lambda_{r}$=6, $\lambda_{r}/\lambda_{l}=1.0$, $\alpha_{xy}^{s}=0.18 \alpha_{0}^{s}$ [$\alpha_{0}^{s}=k_{B}/8\pi$]\cite{Xiao-Q Yu}, $\kappa_{r}=20 (\text{W/mK})$\cite{V. Varshney}, $\theta_{\text{sHl}}$=0.83 and $\sigma_{\text{sHl}}=1.16\pi\times 10^{-2}e^{2}h^{-1}$\cite{W. Feng} are fixed in other three figures. Here, all material parameter are taken for a monolayer MoS$_2$.}
\label{fig2}
\end{figure}

\begin{equation}
\left(
\begin{array}{c}
J_{c} \\
J_{Q}%
\end{array}%
\right) =  G_{\text{H}}\left(
\begin{array}{cc}
1, & \frac{A_{\text{H}}}{G_{H}} \\
\Pi_{\text{H}}, & \frac{K_{\text{H}}}{G_{\text{H}}}+\frac{A_{\text{H}}}{G_{\text{H}}}\Pi_{\text{H}}
\end{array}%
\right)
\left(
\begin{array}{c}
\triangle V \\
-\triangle T
\end{array}
\right),
\label{mmatrix-qc}
\end{equation}
%
$G_{\text{H}}=(J_{c}/\triangle V)_{\triangle T=0}$ is the effective charge conductance of the system, $K_{\text{H}}=-(J_{Q}/\triangle T)_{J_{c}=0}$ is the effective heat conductance for an open electric circuit, $A_{\text{H}}=-(J_{c}/\triangle T)_{\triangle V=0}$ represents nonlocal Nernst conductance, $\Pi_{\text{H}}=(J_{Q}/J_{c})_{\triangle T=0}$ is a nonlocal Peltier coefficient, and $S_{\text{H}}=(\triangle V/\triangle T)_{J_{c}=0}$ denotes a nonlocal Seebeck coefficient of the system. Here, "nonlocal" is used because of the spatial decoupling of the heat current $J_{Q}$ in right arm and charge current $J_{c}$ in left arm. For an ordinary Peltier coefficient and Seebeck coefficient, the four parameters ($J_{Q},J_{c},\triangle T ,\triangle V$) are defined in the same spatial region. Explicit expressions for the various coefficients ($G_{\text{H}}, K_{\text{H}}, A_{\text{H}}, \Pi_{\text{H}}, S_{\text{H}}$) are given in Eq. ({\ref{effective_para}).

%
\section{ results and discussions} \label{results-and-discussions}

\subsection{THE VOLTAGE OUTPUT} \label{results-and-dis A}
In the open circuit case, $J_{c}=0$ and the voltage drop is $V_{\text{open}}=\frac{A_{\text{H}}}{G_{\text{H}}}\triangle T$.
$V_{\text{open}}$ depends on the widths of the arms of the device, as shown in Fig. \ref{fig2}(a) and (b).  A maximum value is attained for a certain range of the geometric parameters (the dark red regions).
In the two limits of $w_{l}\rightarrow0$ or $w_{l}\rightarrow \infty$, $V_{\text{open}}$ tends to zero, as expected. In the latter case, spin coherence is not preserved. 
At a fixed $w_l/\lambda_l$, $V_{\text{open}}$ varies monotonically with $w_{r}/\lambda_{r}$ tending to a constant value (see Fig. {\ref{fig2}}(a)). There is no explicit and severe restriction on the width of right arm ($w_{r}$) for optimizing $V_{\text{open}}$ by only constraining  the ratio of $w_{r}/\lambda_{r}$.

Figs. \ref{fig2}(c) and \ref{fig2}(d) show the variation of $V_{\text{open}}$ with different material quantities. A larger $V_{\text{open}}$ can be obtained by increasing the $\alpha_{xy}^{s}$ of the right arm and the spin Hall angle $\theta_{\text{sHl}}$. Consider now a varying dilute non-magnetic disorder in the left arm, which strongly affects the longitudinal conductivity, while the spin Hall conductivity $\sigma_{\text{sHl}}$ is essentially unchanged (because the spin Hall effect is of topological origin and is protected against such disorder, as long as spin coherence is maintained). Changing the doping thus provides a technologically viable way to optimize the output voltage in the device. The spin diffusion length of the left arm, however, will also be reduced with increasing doping level owing to the decreasing mobility. Thus, one should ensure $w_{l}$ is of the order of spin relaxation length when optimizing the output voltage through doping dilute non-magnetic disorder into the left arm. This can be guaranteed since the lithography resolution can already reach $25$ nm \cite{Sanders}.

On the other hand, the impact of varying the thermal conductivity $\kappa_{r}$ is insignificant (inset in Fig. \ref{fig2}(c)).
We also observe that even in absence of the SNE, there is still non-zero $V_{\text{open}}$ (Fig. \ref{fig2}(c)) which can be ascribed to the combination of SHE and Seebeck effect (SE) (the extra term $\theta_\text{sHr}\sigma_{r}S_{r}$) in Eq. (\ref{msc-current})} in the right arm. The extra term has the following meaning. When a temperature gradient is applied to the right arm, an electric field will be induced along the direction of the temperature gradient owing to the conventional Seebeck effect. The generated electric field would induced a transverse spin current through the SHE, which is superpositioned to the one generated via the SNE. This explains the finite $V_{\text{open}}$ even at zero $\alpha^{s}_{xy}$. Finally, the spin current injected into the left arm induces $V_{\text{open}}$ along the arm direction. From this perspective, the combined effect can be viewed as a generalized SNE.

\begin{figure}[tb]
\centering
\includegraphics[width=1.0\linewidth]{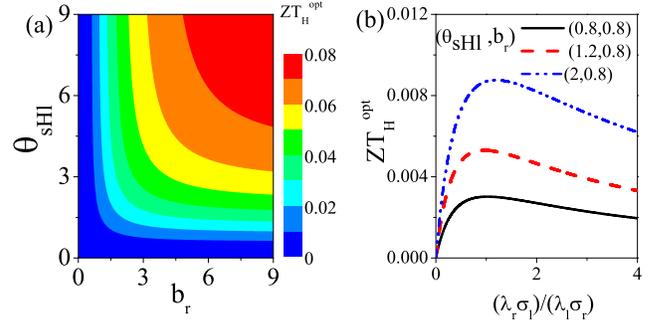}
\caption{(a)  $ZT^{\text{opt}}_{\text{H}}$ as a function of $\theta_{\text{sHl}}$ and $b_{r}$.
(b) $ZT^{\text{opt}}_{\text{H}}$ as function of the ratio $\lambda_{r}\sigma_{l}/\lambda_{l}\sigma_{r}$ for different $\theta_{\text{sHl}}$ and $b_{r}$, for $\theta_{\text{sHr}}=0.15$, and $\frac{L_{1}}{L}=0.8$. In (a) $\lambda_{r}\sigma_{l}/\lambda_{l}\sigma_{r}=1.0$
}
\label{4}
\end{figure}

\subsection{THE FIGURE OF MERIT $ZT_{H}$ OF THE H-SHAPE DEVICE} \label{results-and-dis B}
Fig. \ref{1} (c) shows the equivalent circuit for the proposed device. The output power $P$ of the device is
\begin{equation}
\begin{aligned}
P&=\left(V_{\text{open}}-R_{\text{H}}J_{c}\right)\times J_{c} \\
&=J_{c}\frac{A_{\text{H}}}{G_{\text{H}}}\left\vert \triangle T\right\vert -J_{c}^{2}R_{\text{H}},
\end{aligned}
\end{equation}
where $R_{\text{H}}$ is the internal resistance of the SNE-based device and $R_{\text{H}}J^{2}_{c}$ is the Joule heating produced by the electric current flowing through the internal resistance. Based on Eq. (\ref{mmatrix-qc}), the averaged heat current $J_{Q}$ in the right arm can be given as a function of $J_{c}$
\begin{equation}
\begin{aligned}
J_{Q}&=\frac{A_{\text{H}}}{G_{\text{H}}}TJ_{c}+K_{\text{H}}|\triangle T|. \\
&=e_{\text{H}}TJ_{c}+K_{\text{H}}|\triangle T|.
\end{aligned}
\label{mJQ}
\end{equation}
Compared to the formula for the conventional TE generator (the charge Seebeck effect) \cite{D.Rowe}, the term due to the Joule heating is absent in Eq. (\ref{mJQ}). This makes sense since there is no charge current flowing along the right arm. Thus the power conversion efficiency $\eta_{\text{SNE}}$ can be obtained as a function of $J_{c}$:
\begin{equation}
\eta_{\text{SNE}}(J_{c})= \frac{P}{J_{Q}}
=\frac{J_{c}\frac{A_{\text{H}}}{G_{\text{H}}}\left\vert \triangle T\right\vert -J_{c}^{2}R_{\text{H}}}{\frac{A_{\text{H}}}{G_{\text{H}}}TJ_{c}+K_{\text{H}}|\triangle T|}.
\label{etasne}
\end{equation}
The maximum efficiency is reached at the optimal $J^{\text{opt}}_{c}$, given by
\begin{equation}
J^{\text{opt}}_{c}=\frac{|\triangle T|\frac{A_{\text{H}}}{G_{\text{H}}}}{R_{\text{H}}+R_{\text{load}}^{\text{opt}}}, \ R_{\text{load}}^{\text{opt}}=R_{\text{H}}\sqrt{1+(ZT)_{\text{H}}},
\end{equation}
and has the value
\begin{equation}
\eta^{\text{max} }_{\text{SNE}}=\frac{|\triangle T|}{T}\frac{\sqrt{1+(ZT)_{\text{H}}}-1}{\sqrt{1+(ZT)_{\text{H}}}+1}.
\label{etamax}
\end{equation}
This is a monotonically increasing function of the figure of merit $(ZT)_{\text{H}}$. The ZT value for the present device is
\begin{equation}
(ZT)_{\text{H}}=\frac{(A_{\text{H}})^{2}}{K_{\text{H}}G_{\text{H}}}T=\frac{(S_{\text{H}})^{2}G_{\text{H}}}{K_{\text{H}}}T,
\label{mZTexpress}
\end{equation}
%
where $S_{\text{H}}$ is the effective Seebeck coefficient of the H-shape device. The ZT has a similar expression as that of conventional energy converter.
%
%
Using the explicit expressions for $(ZT)_H$ given in Eq. (\ref{ZT}), we can find the optimal dimensions of the device,
 which are described by the relation of $w_{l}$ and $w_{r}$ and derived from the solutions of the following transcendental equations
\begin{eqnarray}
\cosh \left(\frac{w_l}{\lambda_l}\right)-2\left(\frac{w_l}{\lambda_l}\right)\coth \left(\frac{w_l}{2\lambda_l}\right)&=& 2\theta_{\text{sHl}}^{2}-1, \notag\\
\cosh \left(\frac{w_r}{\lambda_r}\right)-2\left(\frac{w_r}{\lambda_r}\right)\coth \left(\frac{w_r}{2\lambda_r}\right)&=& 2b_{r}^{2}-1,
\label{width_opt}
\end{eqnarray}
where $b_{r}=\Theta^{-1/2}\left(\frac{2e}{\hbar}\sqrt{\frac{{\alpha_{xy}^{s}}^{2}T}{\sigma_{r}\kappa_{r}}}+\sqrt{\frac{S_{r}^{2}
\sigma_{\text{sHr}}^{2}T}{\sigma_{r}\kappa_{r}}}\right)$. The optimal $ZT_{\text{H}}^{\text{opt}}$ of the device can be enhanced by increasing $\theta_{\text{sHl}}=\frac{\sigma_{\text{sHl}}}{\sigma_{l}}$ and $b_{r}$ (Fig. \ref{4}(a)), which can be realized by increasing the parameters ($\alpha_{xy}^{s}$, $\sigma_{\text{sHl}}$) and decreasing the parameters ($\kappa_{r}$, $\sigma_{l}$,).  With $b_{r}$, $\theta_{\text{sHl}}$ and $\theta_{\text{sHr}}$ fixed, there exists an optimal value of the ratio $\frac{\lambda_{r}\sigma_{l}}{\lambda_{l}\sigma_{r}}\approx1$ which yields the largest $ZT_{\text{H}}^{\text{opt}}$ (see Fig. \ref{4}(b)). For the case $\lambda_{r}\simeq\lambda_{l}$, the conductivity of the right arm should be close to that of the left arm to optimize the device.
When examining the ZT value, the present device is not superior to the best traditional devices. The ZT of the proposed device can be larger than 0.008, which is larger than that of a spin Seebeck power generator based on the ISHE (ZT$\sim10^{-4}$ ) \cite{A.B.Cahaya}. With the optimized structure and load resistance, $ZT$ can still be enhanced either by increasing the spin Nernst coefficient of the right arm and spin Hall conductivity of the left arm or by decreasing the charge conductivity and thermal conductivity. It should be mentioned here that the present ZT and that in Ref. \cite{A.B.Cahaya} are both derived in a conventional way by considering the energy conversion from heat to electric power, which differs from the newly proposed spin analog of ZT. 

\section{conclusion}
\label{conclusion}
In summary, the performance of a two-dimensional energy generator based on the concerted effect of the SNE nd ISHE has been studied. It is found that the performance not only depends on the properties of the materials and the geometry, but also on the matching of the load resistance.%
%
It is remarkable that the thermal properties (i.e. thermal conductivity) have little impact on the output voltage. It is interesting to note that, contrary to the conventional TE energy converter, a low charge conductivity enhances the $ZT_{\text{H}}$ here.  This makes it possible to optimize the electrical conductivity, thermal conductivity and Seebeck coefficient simultaneously in a single material. Besides, the heat current in the right arm and the charge current in the left arm are spatially decoupled which excels the conventional TE. The properties of the material in different arms can be manipulated independently. We also speculate that, through the inverse effect (spin Ettingshausen effect), the device can also function as a spin-based thermoelectric refrigerator, when the applied temperature gradient is replaced by an external applied voltage.

\section{acknowledgements}
This work is supported by Hundred Talents Program of The Chinese Academy of Sciences and the NSFC (Grant No. 11674317). G. S. is supported in part by the MOST (Grant Nos. 2012CB932900, 2013CB933401), the NSFC (Grant No. 11474279) and the CAS (Grant No.XDB07010100).  The Center for Nanostructured Graphene (CNG) is sponsored by the Danish National Research Foundation, Project DNRF103.

\appendix

\setcounter{equation}{0}
\setcounter{figure}{0}
\setcounter{table}{0}
\makeatletter
\renewcommand{\theequation}{A\arabic{equation}}
\renewcommand{\thefigure}{A\arabic{figure}}
\renewcommand{\thetable}{A\arabic{table}}

\bigskip
\bigskip

\noindent
\section{Linear transport properties in the right arm} \label{app-ltp}
The linear equation in the right arm of the H-shaped is given in the Eq.(\ref{mt-R-Transport}).
Owing to there is no charge current density in x direction .i.e. $j_{x}^{c}=0 $, therefore, the charge electrochemical potential difference $\partial _{x}\mu _{c}$ in x direction is found to be $-\partial _{x}\mu _{c}/e=\theta _{\text{sHr}}\partial _{y}\mu_{s}/2e+S_{r}\partial _{x}T$
which produces
\begin{equation}
\begin{aligned}
\frac{2e}{\hbar }j_{y}^{s}&=\Theta\sigma_{r}\left(-\frac{\partial _{y}\mu _{s}}{2e}\right)-\left(\theta _{\text{sHr}}\sigma _{r}S_{r}-\frac{2e}{\hbar}\alpha_{xy}^{s}\right)\partial _{x}T ,\\
j_{x}^{Q}&=\left(\theta _{\text{sHr}}S_{r}\sigma _{r}T+\frac{2e}{\hbar}\alpha _{xy}^{s}T\right)\left(\frac{\partial_{y}\mu _{s}}{2e}\right)-{\kappa_{r}}\partial _{x}T,
\end{aligned}
\end{equation}
where  $\Theta=\theta_{\text{sHr}}^{2}+1$. After arrangement, one can get
\begin{equation}
\begin{aligned}
\left(
\begin{array}{c}
\frac{2e}{\hbar }j_{y}^{s} \\
j_{x}^{Q}%
\end{array}%
\right) &=\left(
\begin{array}{cc}
\Theta\sigma _{r} & \Upsilon \\
-\Upsilon T & {\kappa_{r}}%
\end{array}%
\right) 
\left(
\begin{array}{c}
-\partial _{y}\mu _{s}/2e \\
-\partial _{x}T%
\end{array}%
\right).
\end{aligned}
\end{equation}
where $\Upsilon=\theta _{\text{sHr}}\sigma _{r}S_{r}+\frac{2e}{\hbar}\alpha_{xy}^{s}$.
This is the Eq.(\ref{msc-current}) in the main text except here using $\Theta=\theta_{\text{sHr}}^{2}+1$. The spin electrochemical potential $\mu^{s}_{r}$ in y direction obeys the spin-diffusion equation $\partial_{y}^{2}\mu^{s}_{r}=\frac{\mu^{s}_{r}}{\lambda_{r}^{2}}$, which gives $\mu _{r}^{s}=A_{r}e^{\frac{-y}{\lambda _{r}}}+B_{r}e^{\frac{y}{\lambda _{r}}}$
%
where $\lambda_{r}$ is the spin diffusion length of the right arm.  Thus, the heat current $J_{x}^{Q}$ is found to be
\begin{equation}
\begin{aligned}
J_{x}^{Q}&=\int_{-w_{r}}^{0}j_{x}^{Q}dy \\
& =\frac{\zeta_{r}}{\left(e^{\frac{-w_{r}}{\lambda_{r}}}-1\right)}\left(-A_{r}e^{\frac{w_{r}}{\lambda _{r}}}+B_{r}\right)+{\kappa_{r}}w_{r}(-\partial _{x}T),
\end{aligned}
\label{HCurent}
\end{equation}
where $\zeta_{r}=-\frac{\left(\theta _{\text{sHr}}S_{r}\sigma _{r}+\frac{2e}{\hbar}\alpha _{xy}^{s}\right)T}{2e}$. The right arm has been divided into three region $\Omega_{\text{R}1,2,3}$ (see the main text). And the temperature gradient of each region is assumed to be uniform (namely $\nabla^2{T}=0$) and labeled as $\partial_{xi}T$, where $i=1,2,3$ indicate the corresponding region. Hence, one can find
\begin{equation}
\left\{
\begin{aligned}
T_{3}-T_{4}&=\frac{L-L_{1}}{2}\partial _{x1}T, \\
T_{2}-T_{3}&=L_{1}\partial _{x2}T, \\
T_{1}-T_{2}&=\frac{L-L_{1}}{2}\partial _{x3}T,
\end{aligned}
\right.
\end{equation}
where the $T_{4,}T_{3},T_{2},T_{1}$ represents the temperature for x=0,$\frac{L-L_{1}}{2},\frac{L+L_{1}}{2},L$ respectively, besides $\Delta T=T_{1}-T_{4}$ (or =$T_{\text{cold}}-T_{\text{hot}}$) is the temperature difference of the two ends of the right arm. It's intuitively to obtain
\begin{equation}
\Delta T=\frac{L-L_{1}}{2}(\partial _{x1}T+\partial _{x3}T)+L_{1}\partial_{x2}T.
\end{equation}
For bound boundaries in open circuit case, the spin current density conservation at the boundaries $y=0\left(-w_{r}\right)$ gives $j_{y}^{s}(y=-w_{r})=0$ (all regions) and $j_{y}^{s}(y=0)=0$ (regions $\Omega_{\text{R}1,3}$) but $j_{y}^{s}(y=0)=j_{y\text{b}}^{s}$ in region $\Omega_{\text{R}2}$. $j^{s}_{y}$ is an undetermined parameter(the concrete formula will be determined  following) denoting the spin current density of bridge region in y direction.
Thus, we obtain
\begin{equation}
\left\{
\begin{aligned}
&-A_{ri}+B_{ri}=\xi _{r}(-\partial _{xi}T)\text{, \ where }i=1,3 \\
&A_{rj}e^{\frac{w_{r}}{\lambda _{r}}}-B_{rj}e^{-\frac{w_{r}}{\lambda _{r}}}=\xi_{r}\partial _{xj}T,\text{ where }j=1,2,3 \\
&A_{r2}-B_{r2}-\xi _{r}(\partial _{x2}T)=\frac{\lambda _{r}}{\Theta\sigma _{r}}\frac{4e^{2}}{\hbar }j_{y\text{b}}^{s},\\
&\xi _{r}=\frac{\left(\theta _{\text{sHr}}\sigma _{r}S_{r}+\frac{2e}{\hbar}\alpha_{xy}^{s}\right)2\lambda _{r}e}{\Theta\sigma _{r}}.
\end{aligned}
\right.
\label{BC_2}
\end{equation}
Meanwhile, the heat current ($J_{x}^{Q}=\int_{-w_{r}}^{0}j_{x}^{Q}dy)$ conservation at the boundaries $x_{1}=\frac{L-L_{1}}{2}\left(x_{2}=\frac{L+L_{1}}{2}\right)$ giving $J_{x}^{Q}|_{x_{1}^{+}}=J_{x}^{Q}|_{x_{1}^{-}}$ and $J_{x}^{Q}|_{x_{2}^{+}}=J_{x}^{Q}|_{x_{2}^{-}}$. Combing with Eq.(\ref{HCurent}), it yields
\begin{equation}
\begin{aligned}
\begin{smallmatrix}
\left(A_{r1}-A_{r2}\right)\left(1-e^{\frac{w_{r}}{\lambda _{r}}}\right) &+\left(B_{r1}-B_{r2}\right)\left(1-e^{-\frac{w_{r}}{\lambda _{r}}}\right)=\frac{{\kappa_{r}}w_{r}}{\zeta _{r}}\left(\partial_{x2}T-\partial _{x1}T\right), \\
\left(A_{r2}-A_{r3}\right)\left(1-e^{\frac{w_{r}}{\lambda _{r}}}\right) &+\left(B_{r2}-B_{r3}\right)\left(1-e^{-\frac{w_{r}}{\lambda _{r}}}\right)
=\frac{{\kappa_{r}}w_{r}}{\zeta _{r}}\left(\partial_{x3}T-\partial _{x2}T\right).
\end{smallmatrix}
\end{aligned}
\label{BCHeat_1}
\end{equation}

The coefficients $A_{r1}$, $B_{r1}$ , $\partial _{x1}T$ can be proved to be equal to $A_{r3}$, $B_{r3}$ , $\partial _{x3}T$, namely, the spin electrochemical potential distribution and temperature gradient in the region $\Omega _{\text{R1}}$ is equal to that in region $\Omega _{\text{R3}}$. Following is the detail. Based on the Eq. (\ref{BC_2}), we could have
\begin{equation}
\left\{
\begin{array}{c}
A_{r1}=\frac{\xi _{r}}{1+e^{\frac{w_{r}}{\lambda _{r}}}}(\partial _{x1}T)\text{; \
\ \ }B_{r1}=\frac{\xi _{r}}{1+e^{\frac{w_{r}}{\lambda _{r}}}}e^{\frac{w_{r}}{\lambda _{r}}}(-\partial _{x1}T), \\
A_{r3}=\frac{\xi _{r}}{1+e^{\frac{w_{r}}{\lambda _{r}}}}(\partial _{x3}T)\text{; \
\ \ }B_{r3}=\frac{\xi _{r}}{1+e^{\frac{w_{r}}{\lambda _{r}}}}e^{\frac{w_{r}}{\lambda _{r}}}(-\partial _{x3}T).
\end{array}
\right.
\label{AB13}
\end{equation}
The relations in Eq. (\ref{BCHeat_1}) give rise to
\begin{equation}
\begin{smallmatrix}
\left(A_{r1}-A_{r3}\right)\left(1-e^{\frac{w_{r}}{\lambda _{r}}}\right)+\left(B_{r1}-B_{r3}\right)\left(1-e^{-\frac{w_{r}}{\lambda _{r}}}\right)&=\frac{{\kappa_{r}}w_{r}}{\zeta _{r}}\left(\partial_{x3}T-\partial _{x1}T\right).
\end{smallmatrix}
\end{equation}
Taking the $A_{1},B_{1},A_{3},B_{3}$ in Eq. (\ref{AB13}) into the above equation, we get
\begin{equation}
\frac{2\xi _{r}\zeta _{r}(e^{\frac{w_{r}}{\lambda _{r}}}-1)}{(e^{\frac{w_{r}}{\lambda _{r}}}+1)}\left(\partial _{x3}T-\partial _{x1}T\right)={\kappa_{r}}w_{r}\left(\partial_{x3}T-\partial _{x1}T\right).
\end{equation}
Owing to $\frac{2\xi _{r}\zeta _{r}(e^{\frac{w_{r}}{\lambda _{r}}}-1)}{(e^{\frac{w_{r}}{\lambda _{r}}}+1)}\neq {\kappa_{r}}w_{r}$, we can obtain
\begin{equation}
\partial _{x3}T=\partial _{x1}T  \Longrightarrow\left\{
\begin{array}{c}
A_{r1}=A_{r3} \\
B_{r1}=B_{r3}
\end{array}
\right.
\end{equation}
After some algebra, one would obtain six equations with six independent coefficients:
\begin{widetext}
\begin{equation}
\left\{
\begin{aligned}
A_{r1}-\frac{\xi _{r}}{1+e^{\frac{w _{r}}{\lambda _{r}}}}\partial _{x1}T&=0; \quad  B_{r1}+\frac{\xi _{r}}{1+e^{\frac{w _{r}}{\lambda _{r}}}}e^{\frac{w _{r}}{\lambda_{r}}}\partial _{x1}T=0; \quad -e^\frac{w _{r}}{\lambda_{r}}A_{r2}+e^{-\frac{w _{r}}{\lambda_{r}}}B_{r2}=-\xi _{r}\partial _{x2}T ,\\
A_{r2}-B_{r2}-\xi _{r}\left(\partial _{x2}T\right)&=\frac{\lambda _{r}}{\Theta\sigma _{r}}\frac{4e^{2}}{\hbar }j_{y\text{b}}^{s}; \quad \frac{L-L_{1}}{2}\left(\partial _{x1}T+\partial _{x3}T\right)=\Delta T-L_{1}\partial_{x2}T, \\
{\kappa_{r}}w_{r}\left(\partial_{x2}T-\partial _{x1}T\right)&=\zeta _{r}\left[\left(A_{r1}-A_{r2}\right)\left(1-e^{\frac{w_{r}}{\lambda_{r}}}\right)
+\left(B_{r1}-B_{r2}\right)\left(1-e^{\frac{-w _{r}}{\lambda _{r}}}\right)\right].
\end{aligned}
\right.
\end{equation}
\end{widetext}
Finally, we obtain the parameters
\begin{equation}
\begin{aligned}
\partial _{x1}T&=\frac{\Delta T}{L}-\frac{L_{1}P_{r}}{L\xi_{r}};\quad \partial_{x2}T=\frac{\Delta T}{L}+\frac{(L-L_{1})P_{r}}{L\xi _{r}}, \\
A_{r1}&=\frac{\xi _{r}\Delta T}{\left(1+e^{\frac{w_{r}}{\lambda_{r}}}\right)L}-\frac{P_{r}L_{1}}{\left(1+e^{\frac{w_{r}}{\lambda_{r}}}\right)L }, \\
B_{r1}&=-\frac{\xi_{r}\Delta T}{(1+e^{-\frac{w_{r}}{\lambda_{r}}})L}+\frac{P_{r}L_{1}}{\left(1+e^{-\frac{w_{r}}{\lambda_{r}}}\right)L}, \\
A_{r2}&=\frac{\xi _{r}\Delta T}{(1+e^{\frac{w_{r}}{\lambda_{r}}})L}+\frac{2e^{2}}{\hbar}\frac{(1-\text{coth}\frac{w_{r}}{\lambda _{r}})\lambda _{r}}{\Theta_{r}\sigma_{r}}j^{s}_{yb} \\
&+\frac{P_{r}\left(L-L_{1}\right)}{(1+e^{\frac{w_{r}}{\lambda_{r}}})L}, \\
B_{r2}&=-\frac{\xi _{r}\Delta T}{(1+e^{-\frac{w_{r}}{\lambda_{r}}})L}-\frac{2e^{2}}{\hbar}\frac{(1+\text{coth}\frac{w_{r}}{\lambda _{r}})\lambda _{r}}{\Theta\sigma _{r}}j^{s}_{yb} \\
&-\frac{P_{r}\left(L-L_{1}\right)}{(1+e^{\frac{w_{r}}{\lambda_{r}}})L},
\end{aligned}
\end{equation}
where $P_{r}=\frac{4e^{2}}{\hbar }\frac{\lambda _{r}\zeta _{r}\xi_{r}}{\Theta\sigma _{r}\left({\kappa_{r}}w_{r}\text{coth}\frac{w_{r}}{2\lambda_{r}}-2\xi _{r}\zeta_{r}\right )}j_{y\text{b}}^{s}$. Thus, the solutions of the spin-diffusion equation for the region $\Omega_{R1}(\Omega _{R3})$ and $(\Omega _{R1})$ are
\begin{eqnarray}
\mu _{r1}^{s} &=& \mu _{r3}^{s}=-\frac{\xi_{r}\sinh{\frac{w_{r}+2y}{2\lambda _{r}}}}{L\cosh\frac{w_{r}}{\lambda _{r}}}\Delta T+ \frac{4e^{2}}{\hbar }j_{y\text{b}}^{s} \nonumber\\
&\times& \frac{\lambda_{r}\zeta _{r}\xi_{r}L_{1}\left(\sinh \frac{y}{\lambda _{r}}-\sinh\frac{w_{r}+y}{\lambda _{r}}\right)}{\Theta\sigma _{r}\left[\left(1+\cosh\frac{w_{r}}{\lambda _{r}}\right)\kappa_{r}w_{r}-2\xi_{r}\zeta_{r} \sinh\frac{w_{r}}{\lambda _{r}}\right]L}, \nonumber\\
\mu _{r2}^{s} &=& -\frac{\xi_{r}\sinh\frac{w_{r}+2y}{2\lambda _{r}}}{L\cosh \frac{w_{r}}{2\lambda_{r}}} \Delta T-\frac{\lambda _{r}}{\Theta\sigma_{r}}\frac{4e^{2}}{\hbar}j_{y\text{b}}^{s}\left[\frac{\cosh{\frac{w_{r}+y}{\lambda _{r}}}}{\sinh{\frac{w_{r}}{\lambda _{r}}}} \right. \nonumber\\
&-&\left.\frac{\zeta _{r}\xi_{r}\left(\cosh\frac{y}{\lambda _{r}}-\cosh{\frac{w_{r}+y}{\lambda _{r}}}\right)(L-L_{1})}{\left[\left(1+\cosh{\frac{w_{r}}{\lambda _{r}}}\right)\kappa_{r}w_{r}-2\xi_{r}\zeta_{r} \sinh{\frac{w_{r}}{\lambda _{r}}}\right]L}\right].
\label{RSAS}
\end{eqnarray}
Thus
\begin{equation}
\begin{aligned}
\mu^{s}_{r2}|_{y=0}&=-\frac{\xi _{r}\text{tanh}{\frac{w_{r}}{2\lambda_{r}}}}{L}\Delta T+\left[-\frac{\text{coth}{\frac{w_{r}}{\lambda_{r}}}\lambda _{r}}{\Theta\sigma _{r}}+\right.\\
&\left.\frac{\lambda _{r}\zeta _{r}\xi _{r}\left(L-L_{1}\right)\text{tanh}{\frac{w_{r}}{2\lambda_{r}}}}{\Theta\left(-{\kappa_{r}}w_{r}\text{coth}\frac{w_{r}}{2\lambda_{r}}+2\xi _{r}\zeta_{r} \right)\sigma _{r}L}\right]\frac{4e^{2}}{\hbar}j_{y\text{b}}^{s}.
\end{aligned}
\label{RS}
\end{equation}

Taking the $A_{1},B_{1},\partial _{x1}T$ into Eq. (\ref{HCurent}), we can determine the heat current $J_{x}^{Q}$
\begin{widetext}
\begin{equation}
\begin{aligned}
J_{x}^{Q}&=\left(-\kappa_{r}w_{r}+2\xi _{r}\zeta _{r} \tanh\frac{w_{r}}{2\lambda _{r}}\right)\frac{\Delta T}{L}+\frac{4e^{2}}{\hbar }\frac{L_{1}\zeta _{r}\lambda _{r}\tanh\frac{w_{r}}{2\lambda _{r}}}{L\Theta\sigma _{r}}j_{y\text{b}}^{s}.
\end{aligned}
\label{JQ1}
\end{equation}
\end{widetext}

\makeatletter
\renewcommand{\theequation}{B\arabic{equation}}
\renewcommand{\thefigure}{B\arabic{figure}}
\renewcommand{\thetable}{B\arabic{table}}



\section{The transport equation for the left arm in linear-response regime}\label{app-lrr}
When reaching the equilibrium, the charge and spin current densities in the left arm can be written as
\begin{equation}
\left(
\begin{array}{c}
j_{x}^{c} \\
\frac{2e}{\hbar }j_{y}^{s}%
\end{array}%
\right) =\sigma _{l}\left(
\begin{array}{cc}
1 & \theta _{\text{sHl}} \\
-\theta _{\text{sHl}} & 1%
\end{array}%
\right) \left(
\begin{array}{c}
-\partial _{x}\mu^{c}_{l}/e \\
-\partial _{y}\mu^{s}_{l}/2e%
\end{array}%
\right),
\label{matrix_left}
\end{equation}
leading to
\begin{equation}
\left\{
\begin{aligned}
j_{x}^{c}&=-\sigma _{l}\frac{\partial _{x}\mu ^{c}_{l}}{e}-\sigma _{l}\theta_{\text{sHl}}\frac{\partial _{y}\mu^{s}_{l}}{2e}, \\
j_{y}^{s}&=\frac{\hbar }{2e}\sigma _{l}\theta _{\text{sHl}}\frac{\partial _{x}\mu^{c}_{l}}{e}-\frac{\hbar }{2e}\sigma _{l}\frac{\partial _{y}\mu^{s}_{l}}{2e}.
\end{aligned}
\right.
\end{equation}

Similarly, the left arm can be divided into three regions $\Omega_{\text{L}1,2,3}$ as the right arm (detail see the main text). The voltage drop difference in each region is assumed to be uniform, which leads to
\begin{equation}
\left\{
\begin{aligned}
\bigtriangleup V_{1}&=-\frac{L-L_{1}}{2}\left(\frac{\partial _{x}\mu _{l1}^{c}}{e}\right),\\
\bigtriangleup V_{2}&=-L_{1}\left(\frac{\partial _{x}\mu _{l2}^{c}}{e}\right), \\
\bigtriangleup V_{3}&=-\frac{L-L_{1}}{2}\left(\frac{\partial _{x}\mu _{l3}^{c}}{e}\right).
\end{aligned}
\right.
\end{equation}
where the $\bigtriangleup V_{1},\bigtriangleup V_{2}$ and $ \bigtriangleup V_{3}$ represent the voltage drops developed in each corresponding region, respectively. $\bigtriangleup V=V|_{x=0}-V|_{x=L}=\bigtriangleup V_{1}+\bigtriangleup V_{2}+\bigtriangleup V_{3}$ is the total voltage drop
induced in the left arm and is found to be
\begin{equation}
\Delta V=\frac{L_{1}-L}{2e}(\partial _{x}\mu _{l1}^{c}+\partial _{x}\mu_{l3}^{c})-\frac{L_{1}}{e}\partial _{x}\mu_{l2}^{c}.
\end{equation}

Analogously, the spin electrochemical potential $\mu _{li}^{s}(i=1,2,3$ is region index$)$ also obeys the spin diffusion equation ($\partial _{y}^{2}\mu_{li}^{s}=\frac{\mu _{li}^{s}}{\lambda _{l}^{2}}$), which yields
\begin{equation}
\left\{
\begin{aligned}
\mu _{li}^{s}&=A_{li}e^{-\frac{y}{\lambda_{l}}}+B_{li}e^{\frac{y}{\lambda _{l}}},\\
\partial _{y}\mu^{s} _{li}&=\frac{-A_{li}}{\lambda _{l}}e^{-\frac{y}{\lambda_{l}}}+\frac{B_{li}}{\lambda _{l}}e^{\frac{y}{\lambda _{l}}}.
\end{aligned}
\right.
\end{equation}

Similarly, the spin current density conservation at the boundaries $y=w_{l}+d,d$ produce $j_{y}^{s}(y=d+w_{l})=0$ (all regions $\Omega _{l1},\Omega _{l2},\Omega_{l3}$) and $j_{y}^{s}(y=d)=0$ (for regions $\Omega _{l1}$ and $\Omega_{l3}$) but $j_{y}^{s}(y=d)=j_{y\text{b}}^{s}$ (for region $\Omega _{l2}$). Thus we obtain
\begin{equation}
\left\{
\begin{aligned}
2\theta_{\text{sHl}}\lambda_{l}\partial _{x}\mu _{li}^{c}+A_{li}e^{-\frac{w_{l}+d}{\lambda_{l}}}=B_{li}e^{\frac{w_{l}+d}{\lambda_{l}}}, \\
\text{ \ where }i=1,2,3; \\
2\theta_{\text{sHl}}\lambda_{l}\partial _{x}\mu _{li}^{c}+A_{li}e^{-\frac{d}{\lambda_{l}}}=B_{li}e^{\frac{d}{\lambda_{l}}},\\
\text{ \ where }i=1,3; \\
\frac{\theta_{\text{sHl}}\sigma _{l}}{e}\partial _{x}\mu _{l2}^{c}+\frac{\sigma_{l}}{2e\lambda }(A_{l2}e^{-\frac{d}{\lambda_{l}}}-B_{l2}e^{\frac{d}{\lambda_{l}}})&=\frac{2e}{\hbar }j_{y\text{b}}^{s}.
\end{aligned}
\right.
\label{BR1}
\end{equation}

Meanwhile, owing to the charge current ($J_{x}^{c}=\int_{d}^{w_{l}+d}j_{x}^{c}dy) $ conservation at the boundaries $x_{1}$ and $x_{2}$ , we can have $J_{x}^{c}|_{x_{1}^{+}}=J_{x}^{c}|_{x_{1}^{-}}$ and $J_{x}^{c}|_{x_{1}^{+}}=J_{x}^{c}|_{x_{2}^{-}}$. Therefore the charge current is
\begin{equation}
\begin{aligned}
J_{xi}^{c}&=\int_{d}^{w_{l}+d}j_{xi}^{c}dy \\
&=\int_{d}^{w_{l}+d}\left[-\frac{\sigma _{l}}{e}\partial _{x}\mu _{li}^{c}-\frac{\sigma_{l}\theta_{\text{sHl}}}{2e}\partial _{y}\mu _{li}^{s}\right]dy \\
& =-\frac{w_{l}\sigma _{l}}{e}\partial _{x}\mu _{li}^{c}-\frac{\sigma _{l}\theta_{\text{sHl}}}{2e}\int_{d}^{w_{l}+d}\left(\frac{-A_{li}}{\lambda _{l}}e^{-\frac{y}{\lambda_{l}}} \right)\\
&+\left.\frac{B_{li}}{\lambda _{l}}e^{\frac{y}{\lambda _{l}}}\right)dy \\
& =-\frac{w_{l}\sigma _{l}}{e}\partial _{x}\mu _{li}^{c}-\frac{\sigma_{l}\theta_{\text{sHl}}}{2e}\left[A_{li}e^{\frac{-d}{\lambda _{l}}}\left(e^{\frac{-w_{l}}{\lambda_{l}}}-1\right)\right.\\
&\left.+B_{li}e^{\frac{d}{\lambda _{l}}}\left(e^{\frac{w_{l}}{\lambda _{l}}}-1\right)\right],
\end{aligned}
\label{Lcurrent1}
\end{equation}
and
\begin{equation}
\begin{aligned}
&\frac{w_{l}\sigma _{l}}{e}\partial _{x}\mu _{l1}^{c}+\frac{\sigma_{l}\theta_{\text{sHl}}}{2e}\left[A_{l1}EX^{-}+B_{l1}EX^{+}\right] \\
=&\frac{w_{l}\sigma _{l}}{e}\partial _{x}\mu _{l2}^{c}+\frac{\sigma_{l}\theta_{\text{sHl}}}{2e}\left[A_{l2}EX^{-}+B_{l2}EX^{+}\right] \\
=&\frac{w_{l}\sigma _{l}}{e}\partial _{x}\mu _{l3}^{c}+\frac{\sigma_{l}\theta_{\text{sHl}}}{2e}\left[A_{l3}EX^{-}+B_{l3}EX^{+}\right].
\end{aligned}
\label{coeR_1}
\end{equation}
where $EX^{\pm}=e^{\pm\frac{d}{\lambda _{l}}}\left(e^{\pm\frac{w_{l}}{\lambda _{l}}}-1\right)$.

The coefficient $A_{l1}$, $B_{l1}$ , $\partial _{x}\mu _{l1}^{c}$ can be proved to be equal to $A_{l3}$, $B_{l3}$ , $\partial _{x}\mu _{l3}^{c}$. Namely, the spin electrochemical potential distribution and temperature gradient in the region $\Omega _{l1}$ is similar to that in region $\Omega_{l3}$. The details are shown below. According to Eq. (\ref{BR1}) we have
\begin{equation}
\left\{
\begin{aligned}
\frac{\theta_{\text{sHl}}}{e}\partial _{x}\mu _{l1}^{c}+\frac{1}{2e\lambda _{l}}\left(A_{l1}e^{-\frac{w_{l}+d}{\lambda_{l}}}-B_{l1}e^{\frac{w_{l}+d}{\lambda_{l}}}\right)&=0, \\
\frac{\theta_{\text{sHl}}}{e}\partial _{x}\mu _{l1}^{c}+\frac{1}{2e\lambda _{l}}%
\left(A_{l1}e^{-\frac{d}{\lambda_{l}}}-B_{l1}e^{\frac{d}{\lambda_{l}}}\right)&=0, \\
\frac{\theta_{\text{sHl}}}{e}\partial _{x}\mu _{l3}^{c}+\frac{1}{2e\lambda _{l}}%
\left(A_{l3}e^{-\frac{w_{l}+d}{\lambda_{l}}}-B_{l3}e^{\frac{w_{l}+d}{\lambda_{l}}}\right)&=0, \\
\frac{\theta_{\text{sHl}}}{e}\partial _{x}\mu _{l3}^{c}+\frac{1}{2e\lambda _{l}}%
\left(A_{l3}e^{-\frac{d}{\lambda_{l}}}-B_{l3}e^{\frac{d}{\lambda_{l}}}\right)&=0.%
\end{aligned}
\right.
\end{equation}
This leads us to
\begin{equation}
\left\{
\begin{array}{c}
A_{l1}=-\frac{2e^{\frac{w_{l}+d}{\lambda_{l}}}\theta_{\text{sHl}}\lambda _{l}}{1+e^{\frac{w_{l}}{\lambda _{l}}}}\partial _{x}\mu _{l1}^{c}\text{; \ }B_{l1}=\frac{%
2e^{\frac{-d}{\lambda _{l}}}\theta_{\text{sHl}}\lambda _{l}}{1+e^{\frac{w_{l}}{\lambda _{l}}}}\partial _{x}\mu _{l1}^{c}. \\
A_{l3}=-\frac{2e^{\frac{w+d}{\lambda_{l}}}\theta_{\text{sHl}}\lambda _{l}}{%
1+e^{\frac{w_{l}}{\lambda _{l}}}}\partial _{x}\mu _{l3}^{c}\text{; \ }B_{l3}=\frac{%
2e^{\frac{-d}{\lambda _{l}}}\theta_{\text{sHl}}\lambda _{l}}{1+e^{\frac{w_{l}}{\lambda _{l}}}}%
\partial _{x}\mu _{l3}^{c}.
\end{array}%
\right.
\label{Lcoe_13}
\end{equation}
From Eq. (\ref{coeR_1}), we obtain
\begin{equation}
\begin{aligned}
\frac{2w_{l}}{\theta_{\text{sHl}}}(\partial _{x}\mu _{l3}^{c}-\partial _{x}\mu_{l1}^{c})&=(A_{l1}-A_{l3})EX^{-}+(B_{l1}-B_{l3})EX^{+}.
\end{aligned}
\end{equation}
Taking the $A_{l1},B_{l1},A_{l3},B_{l3}$ in Eq. (\ref{Lcoe_13}) into the above equation, which leads
\begin{equation}
\left(\theta_{\text{sHl}}\lambda _{l}\text{cosh}{\frac{d}{\lambda _{l}}}\text{tanh}{\frac{w_{l}}{2\lambda _{l}}}-\frac{w_{l}}{\theta_{\text{sHl}}}\right)\left(\partial _{x}\mu^{c}_{l1}-\partial _{x}\mu ^{c}_{l3}\right)=0.
\end{equation}
Because of the inequality $2 \theta_{\text{sHl}}\lambda _{l}\text{cosh}{\frac{d}{\lambda _{l}}}\text{tanh}{\frac{w_{l}}{2\lambda _{l}}}\neq -\frac{2w_{l}}{\theta_{\text{sHl}}}$, we have
\begin{equation}
\partial _{x}\mu _{l1}^{c}=\partial _{x}\mu _{l3}^{c}=>\left\{
\begin{array}{c}
A_{l1}=A_{l3} \\
B_{l1}=B_{l3}%
\end{array}%
\right.
\end{equation}
After re-arrangement, we obtain six equations with six independent coefficients
\begin{equation}
\begin{aligned}
&\frac{\theta_{\text{sHl}}\sigma _{l}}{e}\partial _{x}\mu_{l2}^{c}+\frac{\sigma_{l}}{2e\lambda_{l}}(A_{l2}e^{-\frac{d}{\lambda_{l}}}
-B_{l2}e^{\frac{d}{\lambda_{l}}})=\frac{2e}{\hbar }j_{y\text{b}}^{s}, \\
&\frac{\theta_{\text{sHl}}}{e}\partial _{x}\mu _{l2}^{c}+\frac{1}{2e\lambda_{l}}(A_{l2}e^{-\frac{w_{l}+d}{\lambda_{l}}}
-B_{l2}e^{\frac{w_{l}+d}{\lambda_{l}}})=0,   \\
&A_{l1}=-\frac{2e^{\frac{w_{l}+d}{\lambda_{l}}}\theta_{\text{sHl}}\lambda _{l}}{1+e^{\frac{w_{l}}{\lambda _{l}}}}\partial _{x}\mu _{l1}^{c};\
\frac{2e^{\frac{-d}{\lambda _{l}}}\theta_{\text{sHl}}\lambda _{l}}{1+e^{\frac{w_{l}}{\lambda_{l}}}}\partial _{x}\mu _{l1}^{c}=B_{l1}, \\
&(L_{1}-L)\partial _{x}\mu _{l1}^{c}-L_{1}\partial _{x}\mu _{l2}^{c}=e\Delta V,\\
&\frac{2w_{l}}{\theta_{\text{sHl}}}(\partial _{x}\mu _{l2}^{c}-\partial _{x}\mu_{l1}^{c})-(A_{l1}-A_{l2})e^{\frac{-d}{\lambda _{l}}}(e^{\frac{-w_{l}}{\lambda_{l}}}-1)=\\
&(B_{l1}-B_{l2})e^{\frac{d}{\lambda _{l}}}(e^{\frac{w_{l}}{\lambda _{l}}}-1).
\end{aligned}
\end{equation}

which produce
\begin{eqnarray}
\partial _{x}\mu _{l1}^{c}&=&-\frac{e\triangle V}{L}-\frac{\frac{2e^{2}}{\hbar }j_{y\text{b}}^{s}\theta_{\text{sHl}}\lambda _{l}L_{1}}{L\left(w_{l}\sigma_{l}\text{coth}{\frac{w_{l}}{2\lambda _{l}}}+2\theta_{\text{sHl}}^{2}\lambda _{l}\sigma _{l}\right)}, \nonumber \\
\partial _{x}\mu _{l2}^{c}&=&-\frac{e\triangle V}{L}+\frac{\frac{2e^{2}}{\hbar }j_{y\text{b}}^{s}\theta_{\text{sHl}}\lambda _{l}\left(L-L_{1}\right)}{L\left(w_{l}\sigma_{l}\text{coth}{\frac{w_{l}}{2\lambda _{l}}}+2\theta_{\text{sHl}}^{2}\lambda _{l}\sigma _{l}\right)}, \nonumber \\
A_{l1}&=&\frac{2e^{\frac{w_{l}+d}{\lambda _{l}}}\theta_{\text{sHl}}\lambda _{l}e}{L(1+e^{\frac{w_{l}}{\lambda _{l}}})}\triangle V+\frac{L_{1}e^{\frac{w_{l}+d}{\lambda _{l}}}P_{l}}{L},\nonumber \\
B_{l1}&=&-\frac{2e^{\frac{-d}{\lambda _{l}}}\theta_{\text{sHl}}\lambda _{l}e}{L\left(1+e^{\frac{w_{l}}{\lambda _{l}}}\right)}\triangle V-\frac{\lambda _{l}^{2}L_{1}e^{\frac{-d}{\lambda _{l}}}P_{l}}{L},\nonumber \\
A_{l2}&=&\frac{2e^{\frac{w_{l}+d}{\lambda _{l}}}\theta_{\text{sHl}}\lambda _{l}e}{L(1+e^{\frac{w_{l}}{\lambda _{l}}})}\triangle V-\frac{\left(L-L_{1}\right)e^{\frac{w_{l}+d}{\lambda _{l}}}P_{l}}{L}\nonumber\\
&&+\frac{ \left(-1+\text{coth}{\frac{w_{l}}{\lambda _{l}}}\right)\lambda _{l}e^{\frac{d+2w_{l}}{\lambda _{l}}}}{\sigma
_{l}}\frac{2e^{2}}{\hbar }j_{y\text{b}}^{s},\nonumber\\
B_{l2}&=&-\frac{2e^{\frac{-d}{\lambda _{l}}}\theta_{\text{sHl}}\lambda _{l}e}{L(1+e^{\frac{w}{\lambda _{l}}})}\triangle V+\frac{\left(L-L_{1}\right)e^{\frac{-d}{\lambda _{l}}}P_{l}}{L}\nonumber \\
&&+\frac{ \left(-1+\text{coth}{\frac{w_{l}}{\lambda _{l}}}\right)\lambda _{l}e^{\frac{-d}{\lambda _{l}}}}{\sigma
_{l}}\frac{2e^{2}}{\hbar }j_{y\text{b}}^{s}, \notag
\end{eqnarray}
where $P_{l}=\frac{4e^{2}}{\hbar}\frac{\theta_{\text{sHl}}^{2}\lambda _{l}^{2}L_{1}}{\left(1+e^{\frac{w_{l}}{\lambda _{l}}}\right)\left(w_{l}\sigma_{l}\text{coth}{\frac{w_{l}}{2\lambda _{l}}}+2\theta_{\text{sHl}}^{2}\lambda _{l}\sigma _{l}\right)}j^{s}_{yb}$.

Owing to $d\ll w_{l}$, here we can approximate $w_{l}+d\approx w_{l}$. The charge current $J_{xi}^{c}$ in the Eq. (\ref{Lcurrent1}) and  spin electrochemical potential $\mu_{l2}^{s}$ are given by, respectively,
\begin{equation}
\begin{aligned}
J_{x}^{c}=&\frac{\Delta V\sigma _{l}}{L}\left(w_{l}+2\theta _{\text{sHl}}^{2}\lambda _{l} \tanh{\frac{w_{l}}{2\lambda _{l}}}\right) \\
&+ \frac{L_{1}}{L}\theta _{\text{sHl}}\lambda _{l}\tanh\left(\frac{w_{l}}{2\lambda _{l}}\right)\frac{2e}{\hbar }j_{y\text{b}}^{s}.
\end{aligned}
\label{JC1}
\end{equation}
\begin{eqnarray}
\mu_{l2}^{s}&=&\frac{2e\sinh\left(\frac{2d+w_{l}-2y}{2\lambda_{l}}\right)\theta_{\text{sHl}}\lambda_{l}}{L\cosh\frac{w_{l}}{2\lambda_{l}}}\Delta V
+\left[\frac{\cosh\frac{d+w_{l}-y}{\lambda_{l}}}{\sinh\frac{w_{l}}{\lambda_{l}}}\right. \nonumber\\
&+&\left.\frac{\left(\cosh{\frac{d-y}{\lambda_{l}}}-\cosh{\frac{d+w_{l}-y}{\lambda_{l}}}\right)(L-L_{l})\theta^{2}_{\text{sHl}}}{\left(\frac{w_{l}}{\lambda_{l}}
+\frac{w_{l}}{\lambda_{l}}\cosh{\frac{w_{l}}{\lambda_{l}}}+2\theta^{2}
_{\text{\text{sHl}}}\sinh{\frac{w_{l}}{\lambda_{l}}}\right)L}
\right] \nonumber\\
&\times & \frac{4e^{2}\lambda_{l}}{\sigma_{l}\hbar}j^{s}_{yb}.
\label{LS}
\end{eqnarray}

\setcounter{equation}{0}
\setcounter{figure}{0}
\setcounter{table}{0}
\makeatletter
\renewcommand{\theequation}{C\arabic{equation}}
\renewcommand{\thefigure}{C\arabic{figure}}
\renewcommand{\thetable}{C\arabic{table}}

\noindent
\section{The spin current density $j_{y\text{b}}^{s}$ in the bridge} \label{app-brigde}

As illustrated in the main text, it's reasonable to assume that the spin current in the bridge regime is uniform and there isn't spin electrochemcial potential accumulation i.e. $\Delta u^{s}=0$, leading to $\mu^{s}|_{y=0}=\mu^{s}|_{y=d}$. Taking the expression of $\mu ^{s}|_{y=0}$ ($\mu^{s}|_{y=d}$) in the Eq.(\ref{RS}) and (\ref{LS}) into this equation, we can determine the spin current density $j_{y\text{b}}^{s}$ as the function of temperature difference $\triangle T$ in right arm and the voltage drop $\triangle V$ in the left arm
\begin{equation}
\begin{aligned}
j_{y\text{b}}^{s}&=-\frac{\hbar}{2e}\left(\frac{\theta_{\text{sHl}}\lambda _{l} \text{tanh}{\frac{w_{l}}{2\lambda _{l}}}}{L}\triangle V+\frac{\xi _{r}\text{tanh}{\frac{w_{r}}{2\lambda _{r}}}}{2Le}\triangle T\right)\\
&/\left[\frac{\lambda _{r}\text{coth}{\frac{w_{r}}{
\lambda _{r}}}}{\Theta\sigma _{r}} +\frac{\lambda _{l}\text{coth}{\frac{w_{l}}{\lambda _{l}}}}{\sigma _{l}}-\frac{\left(L-L_{1}\right)\lambda _{r}}{L\eta _{r}\Theta\sigma _{r}}\right.\\
&\left.-\frac{\left(L-L_{1}\right)\lambda _{l}}{L\tau _{l}\sigma _{l}}\right],
\end{aligned}
\label{spin-birdge}
\end{equation}

where
\begin{equation}
\left\{
\begin{aligned}
\eta _{r}&=\Theta\text{coth}\left(\frac{w_{r}}{2\lambda _{r}}\right)\left(\frac{-{\kappa_{r}}w_{r}\text{coth}{\frac{w_{r}}{2\lambda _{r}}}}{\xi _{r}\zeta _{r}}+2\right), \\
\tau _{l}&=\coth\left({\frac{w_{l}}{2\lambda _{l}}}\right)\left(\frac{w_{l}}{\lambda _{l}}\frac{\coth{\frac{w_{l}}{2\lambda_{l}}}}{\theta_{\text{sHl}}^{2}}+2\right), \\
\xi _{r}\zeta _{r}&=-\frac{(\theta _{\text{sHr}}\sigma _{r}S_{r}+\frac{2e}{\hbar}\alpha_{xy}^{s})^{2}}{\Theta\sigma _{r}}\lambda _{r}T.
\end{aligned}
\right.
\label{coef2}
\end{equation}

For simplicity, we introduce a parameter $\Xi $ as
\begin{eqnarray}
\Xi &=&\frac{\lambda _{r}\text{coth}{\frac{w_{r}}{\lambda _{r}}}}{\Theta\sigma _{r}}+\frac{\lambda _{l}\text{coth}{\frac{w_{l}}{\lambda _{l}}}}{\sigma _{l}}-\frac{(L-L_{1})\lambda _{r}}{\eta_{r}L\sigma _{r}} \notag\\
&-&\frac{(L-L_{1})\lambda _{l}}{\tau _{l}L\sigma _{l}}.
\end{eqnarray}

Hence, spin current $j_{y\text{b}}^{s}$ can be written as
\begin{equation}
\frac{2e}{\hbar}j_{y\text{b}}^{s}=-\frac{\theta_\text{sH1}\lambda _{l}\tanh\frac{w_{l}}{2\lambda _{l}}}{L\Xi}\triangle V-\frac{\xi _{r}\tanh{\frac{w_{r}}{2\lambda _{r}}}}{2Le \Xi }\triangle T.
\label{B-spin-current}
\end{equation}

\setcounter{equation}{0}
\setcounter{figure}{0}
\setcounter{table}{0}
\makeatletter
\renewcommand{\theequation}{D\arabic{equation}}
\renewcommand{\thefigure}{D\arabic{figure}}
\renewcommand{\thetable}{D\arabic{table}}

\noindent
\section{The formulas of figure-of-merit $ZT_{H}$ in the H-shape device} \label{app-zt}
The heat current $J_{Q}$ (i.e. $J^{Q}_{x}$ in Eq.(\ref{JQ1})) in the right arm and charge current $J_{c}$ (namely, $J^{c}_{x}$ in Eq.(\ref{JC1})) in the left arm has been found to be expressed as a linear function temperature difference $\triangle T$ (voltage drop $\triangle V$) in the right (left) arm and spin current density $j_{y\text{b}}^{s}$ in the bridge region , respectively. Whereas $j_{y\text{b}}^{s}$ can be given as linear function of $\triangle T$ and $\triangle V$ in Eq. (\ref{B-spin-current}). Hence, the $J_{Q}$ ($J_{c}$) are also written as the linear function of $\triangle T$ and $\triangle V$
\begin{eqnarray}
J_{Q} &=& \left(\frac{-\kappa_{r}w_{r}+2\xi _{r}\zeta _{r}\tanh{\frac{w_{r}}{2\lambda_{r}}}}{L}-\frac{L_{1}\xi _{r}\zeta _{r}\lambda _{r}\tanh^{2}{\frac{w_{r}}{2\lambda _{r}}}}{L^{2}\Theta\sigma _{r}\Xi }\right) \nonumber\\
&\times& \triangle T+\frac{L_{1}\theta _{\text{sHl}}\lambda _{l}\xi _{r}\tanh{\frac{w_{r}}{2\lambda _{r}}}\tanh{\frac{w_{l}}{2\lambda _{l}}}}{2eL^{2}\Xi }
T\triangle V \nonumber,\\
J_{c}&=& \left(\frac{\sigma _{l}(w_{l}+2\theta _{\text{sHl}}^{2}\lambda _{l}\tanh{\frac{w_{l}}{2\lambda _{l}}})}{L}-\frac{L_{1}\theta _{\text{sHl}}^{2}\lambda _{l}^{2}\tanh^{2}{\frac{w_{l}}{2\lambda _{l}}}}{L^{2}\Xi }\right) \nonumber \\
&\times &\triangle V-\frac{L_{1}\theta_{\text{sH1}}\lambda _{l}\xi _{r}\tanh{\frac{w_{r}}{2\lambda _{r}}}\tanh{\frac{w_{l}}{2\lambda _{l}}}}{2eL^{2}\Xi }\triangle T.%
\label{JQJC}
\end{eqnarray}

Here, we can define the effective charge conductance $G_{\text{H}}=(J_{c}/\triangle V)_{\triangle T=0},$ thermal conductance $K_{H}=-(J_{Q}/\triangle T)_{J_{c}=0}$ and the Peltier coefficient $\Pi_{\text{H}}=(J_{Q}/J_{c})_{\triangle T=0}$, the "Nernst signal" $S_{\text{H}}=(\triangle V/\triangle T)_{J_{c}=0},$ the Nernst conductance $A_{\text{H}}=-(J_{C}/\triangle T)_{\triangle V=0}.$
\begin{equation}
\left\{
\begin{aligned}
G_{\text{H}}&=\frac{\sigma _{1}\left(w_{l}+2\theta_{\text{sHl}}^{2}\lambda _{l}\tanh{\frac{w_{l}}{2\lambda _{l}}}\right)}{L} \\
&-\frac{L_{1}\theta _{\text{sHl}}^{2}\lambda _{l}^{2}\tanh^{2}{\frac{w_{l}}{2\lambda _{l}}}}{L^{2}\Xi }, \\ 
A_{\text{H}}&=\frac{L_{1}\theta _{\text{sH1}}\lambda _{l}\xi _{r}\tanh{\frac{w_{r}}{2\lambda _{r}}}\tanh{\frac{w_{l}}{2\lambda _{l}}}}{2eL^{2}\Xi }, \\
\Pi _{\text{H}}&=\frac{A_{\text{H}}}{G_{\text{H}}}T , \ S_{\text{H}}=\frac{A_{\text{H}}}{G_{\text{H}}}, \\ 
K_{\text{H}}&= \frac{\kappa_{r}w_{r}-2\xi _{r}\zeta _{r}\tanh{\frac{w_{r}}{2\lambda _{r}}}}{L}+\frac{L_{1}\xi _{r}\zeta _{r}\lambda _{r}\tanh^{2}{\frac{w_{r}}{2\lambda _{r}}}}{L^{2}\left(\theta^{2}_{\text{sHr}}+1\right)\sigma
_{r}\Xi }\\
&-\frac{(A_{\text{H}})^{2}T}{G_{\text{H}}}.
\end{aligned}
\right.
\label{effective_para}
\end{equation}

From Eq. (\ref{JQJC}) and (\ref{effective_para}), one can obtain
\begin{equation}
\begin{aligned}
\left(
\begin{array}{c}
J_{c} \\
J_{Q}%
\end{array}%
\right)
& =  G_{\text{H}}\left(
\begin{array}{cc}
1, & \frac{A_{\text{H}}}{G_{H}} \\
\Pi_{\text{H}}, & \frac{K_{\text{H}}}{G_{\text{H}}}+\frac{A_{\text{H}}}{G_{\text{H}}}\Pi_{\text{H}}
\end{array}%
\right)
\left(
\begin{array}{c}
\triangle V \\
-\triangle T
\end{array}
\right) \\
&=  G_{\text{H}}\left(
\begin{array}{cc}
1, & \frac{A_{\text{H}}}{G_{H}} \\
\frac{A_{\text{H}}T}{G_{\text{H}}}, & \frac{K_{\text{H}}}{G_{\text{H}}}+\frac{A_{\text{H}}}{G_{\text{H}}}\frac{A_{\text{H}}T}{G_{\text{H}}}
\end{array}%
\right)
\left(
\begin{array}{c}
\triangle V \\
-\triangle T
\end{array}
\right),
\end{aligned}
\label{matrix-qc}
\end{equation}

Thus, the open voltage $V_{\text{open}}$ (namely, the charge current $J_{c}=0$) is found to be
\begin{widetext}
\begin{equation}
\begin{aligned}
V_{\text{open}}=\frac{A_{H}}{G_{H}}\triangle T=\frac{L_{1}\theta
_{\text{sH1}}\xi _{r}\text{tanh}{\frac{w_{r}}{2\lambda _{r}}}\text{tanh}{\frac{w_{l}}{%
2\lambda _{l}}}\triangle T}{2\sigma _{1}\left(\frac{w_{l}}{\lambda _{l}}+2\theta_{\text{sHl}}^{2}\text{tanh}{\frac{w_{l}}{2\lambda _{l}}}\right)\Xi eL-2L_{1}\theta_{\text{sHl}}^{2}\lambda_{l}^{{}}\text{tanh}^{2}{\frac{w_{l}}{2\lambda _{l}}}e},
\end{aligned}
\end{equation}
\end{widetext}

Here, we introduce dimensionless coefficient $\Xi^{\prime }=\Xi\frac{\sigma _{1}}{\lambda _{l}}$ and take the formulas of $\xi _{r}$ into Eq. (\ref{coef2}), we can obtain

\begin{equation}
\begin{aligned}
V_{\text{open}}&=\frac{A_{\text{H}}}{G_{\text{H}}}\triangle T \\
&=\frac{\theta _{\text{sH1}}\tanh{\frac{w_{r}}{2\lambda _{r}}}\tanh{\frac{w_{l}}{2\lambda _{l}}}\left(\theta _{\text{sHr}}S_{r}+\frac{\frac{2e}{\hbar}\alpha _{xy}^{s}}{\sigma _{r}}\right)}{\left(\frac{w_{l}}{\lambda _{l}}+2\theta _{\text{sHl}}^{2}\tanh{\frac{w_{l}}{2\lambda _{l}}}\right)\Xi ^{\prime }\frac{L}{L_{1}}-\theta _{\text{sHl}}^{2}\tanh^{2}{\frac{w_{l}}{2\lambda _{l}}}}\\
&\times\frac{\lambda _{r}^{{}}}{\lambda _{l}}\frac{\triangle T}{\Theta}.
\end{aligned}
\end{equation}

The induced voltage in left arm by the temperature difference $\triangle T$ via combination of the spin Nernst effect and inverse spin Hall effect is $\frac{A_{\text{H}}}{G_{\text{H}}}\triangle T$ ($\triangle T=T_{\text{cold}}-T_{\text{hot}},|\triangle T|=-\triangle T$), Therefore, the voltage drop on the load (output voltage) is found to be
\begin{equation}
\begin{aligned}
V&=\frac{A_{\text{H}}}{G_{\text{H}}}|\triangle T|-J_{c}R_{\text{H}}, \\
\end{aligned}
\end{equation}

The output power $W$ can then be represented as a function of $J_{c}$
\begin{equation}
W=VJ_{c}=J_{c}\frac{A_{\text{H}}}{G_{H}}\left\vert \triangle T\right\vert -J_{c}^{2}R_{\text{H}},
\end{equation}

From the Eq.(\ref{matrix-qc}), we can have
\begin{equation}
\triangle V=\frac{J_{c}}{G_{\text{H}}}+\frac{A_{\text{H}}}{G_{\text{H}}}\triangle T,
\end{equation}

Giving that
\begin{equation}
\begin{aligned}
J_{Q}&=A_{\text{H}}T\triangle V-\left(K_{\text{H}}+\frac{A_{\text{H}}A_{\text{H}}}{G_{\text{H}}}T\right)\triangle T \\
&=A_{\text{H}}T(\frac{J_{c}}{G_{\text{H}}}+\frac{A_{\text{H}}}{G_{\text{H}}}\triangle T)-(K_{\text{H}}+\frac{A_{\text{H}}A_{\text{H}}}{G_{\text{H}}}T)\triangle T \\
&=\frac{A_{\text{H}}}{G_{\text{H}}}TJ_{c}-K_{\text{H}}\triangle T    \\
&=\frac{A_{\text{H}}}{G_{\text{H}}}TJ_{c}+K_{\text{H}}|\triangle T|.
\end{aligned}
\end{equation}

The power conversion efficiency can also be given as a function of $J_{c}$
\begin{equation}
\begin{aligned}
\eta (J_{c})=\frac{W}{J_{Q}}
=\frac{J_{c}\frac{A_{\text{H}}}{G_{\text{H}}}\left\vert \triangle T\right\vert -J_{c}^{2}R_{\text{H}}}{\frac{A_{\text{H}}}{G_{\text{H}}}TJ_{c}+K_{\text{H}}|\triangle T|}.
\end{aligned}
\end{equation}

The maximum efficiency of this power conversion scheme $\eta _{\max }^{\text{SNE}}$ is reach at the optimal $J^{\text{opt}}_{c}$
\begin{equation}
J^{\text{opt}}_{c}=\frac{|\triangle T|\frac{A_{\text{H}}}{G_{\text{H}}}}{R_{\text{H}}+R_{\text{load}}^{\text{op}}}, \ R_{\text{load}}^{\text{opt}}=R_{\text{H}}\sqrt{1+(ZT)_{\text{H}}}.
\end{equation}

Thus
\begin{equation}
\begin{aligned}
\eta _{\max }^{\text{SNE}}& =\frac{|\triangle
T|}{T}\frac{2+(ZT)_{\text{H}}-2\sqrt{1+(ZT)_{\text{H}}}}{(ZT)_{\text{H}}} \\
&=\frac{|\triangle T|}{T}\frac{\sqrt{1+(ZT)_{\text{H}}}-1}{\sqrt{1+(ZT)_{\text{H}}}+1}.
\end{aligned}
\end{equation}

The value of the spin Nernst figure of merit for the ISHE scheme is
\begin{equation}
\begin{aligned}
(ZT)_{\text{H}}
=\frac{(A_{\text{H}})^{2}R_{\text{H}}}{K_{\text{H}}}T
=\frac{(S_{\text{H}})^{2}}{K_{\text{H}}R_{\text{H}}}T,
\end{aligned}
\end{equation}

Taking the expression of $A_{\text{H}},R_{\text{H}},K_{\text{H}}$ in Eq.(\ref{effective_para}) into it, we can determine the ZT value of the H-shape system
\begin{widetext}
\begin{eqnarray}
(ZT)_{\text{H}}&=&\frac{1}{m-1}, \nonumber\\
\text{where}\quad m &=&\left[-1+\frac{L}{L_{1}}\Xi ^{\prime }\coth\left({\frac{w_{l}}{2\lambda _{l}}}\right)\left(\frac{w_{l}}{\lambda _{l}}\frac{\coth{\frac{w_{l}}{2\lambda _{l}}}}{\theta_{\text{sHl}}^{2}}+2\right)\right]\times  \left[-1+\frac{L\Xi ^{\prime \prime}}{L_{1}}{\Theta}\coth\left({\frac{w_{r}}{2\lambda _{r}}}\right)\left(\frac{w_{r}}{\lambda _{r}}\frac{\coth{\frac{w_{r}}{2\lambda _{r}}}}{b_{r}}+2\right)\right] \nonumber\\
\label{ZT}
\end{eqnarray}
\end{widetext}
Where $\Xi^{\prime }=\Xi \frac{\sigma _{1}}{\lambda _{l}}, \Xi ^{\prime \prime }=\Xi \frac{\sigma _{r}}{\lambda _{r}}$.

\setcounter{equation}{0}
\setcounter{figure}{0}
\setcounter{table}{0}
\makeatletter
\renewcommand{\theequation}{E\arabic{equation}}
\renewcommand{\thefigure}{E\arabic{figure}}
\renewcommand{\thetable}{E\arabic{table}}
\section{Some comments on the H-shape device} \label{app-sm}
Comments on applying the temperature gradient to the right arm:
\begin{enumerate}
\item	In our conceptual study, the temperature gradient is assumed to exist only in one arm of the H-shape device. We believe this can be achieved in experiments. For example, heater coils and laser beams have been used in experiments. For the latter, the size and positon of the laser spots can be controlled precisely in experiments: the laser spots can be positioned between contacts which are about 1 $\mu$m away from each other \cite{Buscema}. Therefore, it should be possible to control the position of the laser beam to the right arm of the H-shape detector.
\item	Furthermore, one arm of the H-shape detector can be made longer with a larger-sized pad so that the laser spot can be easily applied to the pad.
\end{enumerate}

Comments on having the temperature gradient in the two arms simultaneously:\\

If the left arm unintentionally experiences a temperature gradient, the Seebeck effect may cause an even larger voltage drop at the two ends of the left arm. Moreover the temperature gradient on the left arm is in the same direction to that in the right arm; it induces a transverse spin current in the same direction as that induced by the right-arm temperature gradient. Therefore, the left gradient does not cancel the effect due to the right temperature gradient, but instead enhances the total output. To make the discussion simple and clear, we only consider the situation where the temperature is applied to the right arm.


\end{document}